\documentstyle[12pt]{article}
\newcommand{\Lambdabm}{\mbox{\boldmath $\Lambda$}}
\newcommand{\taubm}{\mbox{\boldmath $\tau$}}
\newcommand{\pibm}{\mbox{\boldmath $\pi$}}
\newcommand{\rbm}{\mbox{\boldmath $r$}}
\newcommand{\xbm}{\mbox{\boldmath $x$}}
\newcommand{\Rbm}{\mbox{\boldmath $R$}}
\makeatletter
\def\lsim{\mathrel{\mathpalette\gl@align<}}
\def\gsim{\mathrel{\mathpalette\gl@align>}}
\def\gl@align#1#2{\lower.6ex\vbox{\baselineskip\z@skip\lineskip\z@
    \ialign{$\m@th#1\hfil##\hfil$\crcr#2\crcr\sim\crcr}}}
\makeatother
\input{epsf}
\begin{document}
\title{H-dibaryon matter in the Skyrme model on a hypersphere}
\author{T. Sakai\thanks{E-mail: tsakai@rcnp.osaka-u.ac.jp}\, and 
H. Suganuma \\ 
RCNP, Osaka University, Ibaraki, Osaka 567-0047, Japan}

\maketitle

\begin{abstract}
We study the properties of H-dibaryon matter through the SO(3) Skyrmion
solution on a three-dimensional hypersphere $S^3$.
As the density increases, the swelling of H-dibaryon is found.
Above a critical density, the system becomes uniform in terms of the baryon 
density.
In this uniform phase, the critical order parameter is largely reduced, which 
can be interpreted as the chiral symmetry restoration.
From the comparison with the SO(2)$\times$SO(2) solution, the SO(3) soliton 
is found to be the true ground state for high density system with 
$R \leq 0.64$ fm in the $B=2$ sector with $N_{\rm f}=3$.
\end{abstract}

\section{Introduction}
Recently exotic hadron states have been studied elaborately.
Among them, an H-dibaryon has been an object of many works\cite{Sakai95} 
as a hopeful candidate of $q^6$ state since the prediction by 
Jaffe\cite{Jaffe77} in the MIT bag model.
Lots of theoretical studies have been made on the problem of the 
H-dibaryon, and have predicted the H-dibaryon mass below the 
$\Lambda\Lambda$ threshold\cite{Sakai95,Golowich92}.
The Skyrme model was first applied to the H-dibaryon by Balachandran 
{\it et al.} proposing the $B=2$ SO(3) soliton configuration, which has the 
symmetric configuration in terms of the flavor 
SU(3)\cite{Balachandran84,Jaffe85}.
They obtained the H-dibaryon mass lighter than two $\Lambda$'s ($M_{\rm H} 
\sim 1.92 M_{\rm B}$) in the SU(3) flavor symmetric limit.

Experimentally, several candidate events have been 
reported\cite{Shahbazian88,Longacre95}, but any conclusive result for its 
existence is not obtained yet.
The binding energy of two $\Lambda$'s in a double hypernucleus places a 
restriction on the H-mass.
A double hypernucleus event found at KEK\cite{Aoki90} implies that the lower 
limit of the H-dibaryon mass is a few tens of MeV below the $\Lambda\Lambda$ 
threshold.

The possibility of the existence of H-dibaryon matter in a high-density neutron 
star is pointed out by Tamagaki {\it et al.}\cite{Tamagaki91}.
In their argument, the compactness of the H-dibaryon radius is an important 
assumption on the existence of H-dibaryon matter, although the size of 
H-dibaryon may vary in the dense system as the nucleon swelling phenomenon 
in nuclear matter\cite{Ichie95}.
If so, nuclear matter may decay into H-matter at a critical density 
$\rho_{\rm c} = 6 \sim 9 \rho_0$ ($\rho_0 = 0.17$ fm$^{-3}$) before the 
phase transition into quark-gluon plasma.

The properties of nuclear matter at high densities have been investigated 
in many ways.
Castillejo {\it et al.} studied Skyrmion matter by placing SU(2) hedgehog 
Skyrmions on a lattice and show that the deconfinement phase transition and 
the chiral symmetry restoration occur above a critical 
density\cite{Castillejo89}.
Manton and Ruback obtained the similar results by a simple and 
mathematically sophisticated approach.
Instead of placing many Skyrmions in the flat space $\Rbm ^3$, Skyrmion 
matter is approximated by one Skyrmion on an $S^3$ hypersphere with radius $R$.
They showed that beyond a critical density the lowest energy solution is given 
by the ``identity map" and the baryon density distributes uniformly on the 
hypersphere\cite{Manton86}.
In this phase, the spacially averaged value of the chiral order parameter 
becomes zero, which may imply the chiral symmetry restoration at high 
densities.
This phase transition was investigated also in the case of two Skyrmions on 
$S^3$\cite{Jackson87}.
The result is qualitatively the same as one Skyrmion case on $S^3$, and they 
concluded that these behaviors are a general tendency in the Skyrme model.

In this paper, we investigate the feature of H-dibaryon matter by using the 
SO(3) Skyrmion with baryon number 2 located on $S^3$ hypersphere.
We study how the QCD phase transition occurs in this situation.
We study also how the size of the H-dibaryon varies in high-density H-matter.
For simplicity on the mathematical and physical insight, we deal with the 
chiral limit in this paper.

\section{Formalism}
The Skyrme Lagrangian in the chiral limit\cite{Balachandran84,Zahed86} is 
given by
\begin{equation}
{\cal L} = -\frac{f_{\pi}^2}{2} {\rm Tr} (\partial_{\mu} U^{\dagger} 
\partial^{\mu} U ) + \frac{1}{32 e^2} {\rm Tr} [ \partial_{\mu} U 
U^{\dagger} , \partial_{\nu} U U^{\dagger} ]^2 + {\cal L}_{\rm WZ} ,
\end{equation}
where $f_{\pi}$ and $e$ denote the pion decay constant and the Skyrme
parameter, respectively.
Here, 
$U(\xbm) \equiv e^{i\frac{\lambda ^a}{2} \pi^a (\xbm) / f_{\pi} } \in 
{\rm SU(3)}_{\rm f}$
is described by the Nambu-Goldstone mode, $\pi^a (\xbm)$ ($a = 1, \cdots , 8$), 
and ${\cal L}_{\rm WZ}$ denotes the Wess-Zumino term.
The ansatz for the chiral field of SO(3) soliton on $S^3$ is represented 
with two profile functions, $\chi(\mu)$ and $\psi(\mu)$, 
as\cite{Balachandran84,Jaffe85}
\begin{equation}
U = \exp \left[ i \left\{ \Lambdabm \cdot \hat{\rbm} \chi (\mu) - \left( 
\frac{3}{2} ( \Lambdabm \cdot \hat{\rbm} )^2 -1 \right) \psi (\mu) \right\} 
\right] .
\end{equation}
Here, $\Lambdabm = (\Lambda_1, \Lambda_2, \Lambda_3) = (\lambda_7, -\lambda_5, 
\lambda_2)$ is the generators of the SO(3) group, where $\lambda_a$
($a=1,2,\cdots,8$) are Gell-Mann matrices, and three polar 
coordinates on $S^3$ are $\mu$, $\theta$, and $\phi$.
The baryon current density $B^{\nu} (\mu)$ and the energy of the Skyrmion 
$E(U)$ on $S^3$ with radius $R$ can be easily obtained from those on 
$\Rbm^3$\cite{Balachandran84} by simple replacement\cite{Makhankov93}, 
\begin{eqnarray}
& & r \rightarrow R \sin \mu ,
\nonumber \\
& & {\rm d}r \rightarrow R {\rm d}\mu ,
\\
& & {\rm d}V = 4\pi r^2 {\rm d}r \rightarrow 
{\rm d}V = 4\pi R^3 \sin^2 \mu {\rm d}\mu .
\nonumber
\end{eqnarray} 
The winding number, which is identified with the baryon number, is 
\begin{equation}
B = 4\pi R^3 \int_0^{\pi} \sin^2 \mu {\rm d}\mu B^0 (\mu) ,
\end{equation}
where $B^0 (\mu)$ is the baryon number density,
\begin{equation}
B^0 (\mu) = \frac{\epsilon^{0\alpha\beta\gamma}}{24\pi^2} {\rm Tr} 
[ L_{\alpha} L_{\beta} L_{\gamma} ] 
= -\frac{1}{2\pi^2 R^3 \sin^2 \mu} \left[ 
\frac{{\rm d}\chi}{{\rm d}\mu} (1- \cos \chi \cos \frac{3}{2} \psi ) 
+ \frac{3}{2} \frac{{\rm d}\psi}{{\rm d}\mu} \sin \chi \sin \frac{3}{2} \psi 
\right] ,
\end{equation}
with $L_{\alpha} = U^{\dagger} \partial_{\alpha} U$.
For the $B=2$ Skyrmion on $S^3$, the profile functions satisfy the boundary 
conditions\cite{Balachandran84},
\begin{equation}
\chi (0) = \pi , \qquad \psi (0) = \frac{2}{3} \pi, \qquad
\chi (\pi) = 0 , \qquad \psi (\pi) = 0 .
\label{boundcond}
\end{equation}

The static energy of the SO(3) Skyrmion on $S^3$ is written as
\begin{eqnarray}
E(U) &=& \int {\rm d}V \left[ \frac{f_{\pi}^2}{2} {\rm Tr} (\partial_{\mu} 
U^{\dagger} \partial^{\mu} U ) - \frac{1}{32 e^2} {\rm Tr} [ \partial_{\mu} U 
U^{\dagger} , \partial_{\nu} U U^{\dagger} ]^2 \right]
\nonumber \\
     &\equiv& f_{\pi}^2 R I_1 + \frac{1}{e^2 R} I_2 ,
\end{eqnarray}
where $I_1$ and $I_2$ are dimensionless integrals,
\begin{eqnarray}
I_1 &=& 4\pi \int_0^{\pi} {\rm d}\mu \sin^2 \mu \left[ 
\left( \frac{{\rm d}\chi}{{\rm d}\mu} \right)^2 +
\frac{3}{4} \left( \frac{{\rm d}\psi}{{\rm d}\mu} \right)^2 + 
\frac{4}{\sin^2 \mu} \left\{ 1-\cos \chi \cos \frac{3}{2} \psi \right\} 
\right] ,
\\
I_2 &=& 2\pi \int_0^{\pi} {\rm d}\mu \left[ \frac{1}{\sin^2 \mu} \left\{ 
3 \sin^2 \chi \sin^2 \frac{3}{2} \psi + \left( 1- \cos \chi \cos \frac{3}{2} 
\psi \right)^2 \right\} \right.
\nonumber \\
& & \hspace{15mm} + \left( 1- \cos \chi \cos \frac{3}{2} \psi \right) 
\left\{ \left( \frac{{\rm d}\chi}{{\rm d}\mu} \right)^2 +
\frac{9}{4} \left( \frac{{\rm d}\psi}{{\rm d}\mu} \right)^2 \right\}  
\left. + 3 \sin \chi \sin \frac{3}{2} \psi \, 
\frac{{\rm d}\chi}{{\rm d}\mu} \, \frac{{\rm d}\psi}{{\rm d}\mu} \right] .
\end{eqnarray}
The Euler-Lagrange equations for $\chi$ and $\psi$ are obtained by extremizing 
the static energy with respect to them.

\section{Numerical results and concluding remarks}
The profile functions are obtained numerically by extremizing the energy 
functional $E(U)$ under the boundary conditions (\ref{boundcond}).
We show in Fig.1 the profile functions of the $B=2$ SO(3) Skyrmion on $S^3$ 
in typical two cases using the 
parameters in ref.\cite{Balachandran84}: $f_{\pi} = 67$ MeV and $e = 4.47$.
(In Refs.\cite{Manton86,Jackson87}, the energy scale is rescaled by 
$f_{\pi}/(\sqrt{2}e) = 10.6$ MeV, and the length scale by  
$1/(\sqrt{2}ef_{\pi}) = 0.466$ fm to reduce variable dimensionless.)
There are always two physically equivalent solutions: one is localized 
around the north pole, and another is localized around the south pole when 
$R$ is large.
The two solutions, $(\chi_1, \psi_1)$ and $(\chi_2, \psi_2)$, have the same 
energy, and have the relation,
\begin{eqnarray}
\chi_2 (\mu) &=& \pi - \chi_1 (\pi - \mu),
\nonumber \\
\psi_2 (\mu) &=& \frac{2}{3} \pi - \psi_1 (\pi - \mu) .
\end{eqnarray}
At low densities, the Skyrmion localizes at either pole.
In Fig.1(a), the profile functions for $R = 2$ fm are shown, where the
Skyrmion is localized around the south pole.
The behavior of the profile functions is similar to that for the Skyrmion on 
the flat space $\Rbm^3$\cite{Jaffe85}.
On the other hand, as shown in Fig.1(b), the profile functions spread all over 
the hypersphere at high densities.

Fig.2 shows the energy of the SO(3) Skyrmion solution with the lowest energy 
as a function of the hypersphere radius $R$.
At large $R$, the energy of the Skyrmion on $S^3$ approaches the one on
$\Rbm ^3$.
The behavior of the energy is essentially parallel with the SU(2) 
Skyrmion on $S^3$\cite{Manton86}.
The character of the lowest energy solution changes around $R\sim 1$ fm.
The minimum energy is attained at $R \simeq 0.57$ fm, and 
the minimum value is only 4 \% above the BPS saturation value 
$E_{\rm BPS} \equiv 6\sqrt{2}\pi^2 (f_{\pi}/e) |B|$ (Bogomol'ny bound).
It can be shown that the BPS saturation\cite{Zahed86} is never attained for 
the SO(3) Skyrmion on $S^3$ unlike the SU(2) Skyrmion on $S^3$.
In the $B=1$ SU(2) case, the BPS saturation is achieved by the ``identity map",
$U(\mu, \theta, \varphi ) = \cos f(\mu) + i \taubm \cdot \hat{\rbm} \sin 
f(\mu)$ with $f(\mu) = \mu$ or $f(\mu) = \pi - \mu$ \cite{Manton86}.
The SU(2) solution on $S^3$ becomes ``identity map" for small $S^3$ as 
$R \leq R_{\rm c}$.
On the other hand, in the SO(3) soliton solution, there are two kinds of 
profile functions, $\chi (\mu)$ and $\psi (\mu)$, and the $B=2$ SO(3) solution 
on $S^3$ tends to approach $\chi (\mu) = \frac{3}{2} \psi (\mu) = \pi - \mu$ 
as $R$ decreases as shown in Figs.1(a) and (b).
Moreover, there is a similarity in that the profile function has linear
dependence on $\mu$, so that we call the $B=2$ SO(3) solution with 
$\chi (\mu) = \frac{3}{2} \psi (\mu) = \pi - \mu$ as the ``identity map" in 
this paper.
Being different from the SU(2) case, the ``identity map" does not satisfy 
the Euler-Lagrange equations so that it is slightly above the lowest energy
curve at high densities.

At the low density phase, the baryon density is localized around either the 
north or the south pole, while at the high density phase the SO(3) Skyrmion 
spreads over the hypersphere, and loses its identity as a soliton.
This may correspond to a phase transition from the confinement phase to the 
deconfinement phase.
In Fig.3, we show the baryon number densities $B^0(\mu)$ of the solutions of 
several hypersphere radii around this phase transition.
We can clearly see the localized-delocalized transition.

The phase transition occurs so smoothly that it is difficult to find the 
critical density strictly.
If we put the critical radius $R_{\rm c}$ to be 0.8$\sim$1.0 fm, the 
corresponding critical density $\rho_{\rm c} = 2/(2\pi^2 R_{\rm c}^3)$ is 
0.1$\sim$0.2 baryons/fm$^3$, which is the same order of magnitude as the 
equilibrium nuclear matter density $\rho_0$.
This is rather smaller value than the expected critical density which is the 
order of several times $\rho_0$.
Too low critical density problem appears also in $B=1$ hedgehog Skyrmion
matter\cite{Castillejo89,Manton86,Jackson87}.
The critical density $\rho_{\rm c}$ is 0.07 $\sim$ 0.18 fm$^{-3}$ in these 
models.
One of the reason is probably the lack of the kinetic energy as stated in 
ref.\cite{Jackson87}.
Nevertheless, it is interesting to be successful in modeling the high density 
phase transition in QCD qualitatively by such simple models.

The order parameter $\sigma_0$, which is a measure of the chiral symmetry
breaking, is introduced by the spacial average\cite{Forkel89},
\begin{equation}
\sigma_0 \equiv \left| \int_{S^3} {\rm d}V \frac{1}{3} {\rm Tr} U \right| 
= \frac{2}{\pi} \left| \int_0^{\pi} \sin^2 \mu {\rm d}\mu \left( \frac{1}{3} 
e^{i \psi} + \frac{2}{3} \cos \chi e^{-\frac{i}{2}\psi} \right) \right| .
\label{chiralorderparam}
\end{equation}
As shown in Fig.4, the order parameter $\sigma_0$ becomes smaller at 
high densities, which is considered to be the sign of the chiral symmetry
restoration.
The non-zero value of the order parameter may be due to the limitation of the 
non-linear sigma model.

The Skyrmion root mean square (r.m.s.) radius is given by
\begin{equation}
\bar{r} \equiv \sqrt{\langle ( R \sin\mu)^2 \rangle} = \sqrt{\left. \int_{S^3} 
{\rm d}V B^0(\mu) (R\sin\mu)^2 \right/ 2 }
\label{rmsradius}
\end{equation}
and shown in Fig.5.
The contributions from the upper half ($0 \leq \mu \leq \pi /2$) and the lower 
half ($\pi /2 \leq \mu \leq \pi$) of the hypersphere are also shown.
At low densities, $\bar{r}$ is almost determined by the contribution from the 
half of the hypersphere where the Skyrmion localizes, while the both 
contributions are almost the same at high densities.
Thus the deconfinement phase transition is read also in Fig.5.
At low densities, the Skyrmion size coincides with the size of the Skyrmion 
on $\Rbm^3$.
In the localized phase ($R \geq R_{\rm c} \simeq 0.8 - 1.0$ fm), the size 
$\bar{r}$ of the SO(3) Skyrmion becomes larger corresponding to the spread 
of its profile functions for small $R$ or at high densities.
This enlargement implies the swelling of the H-dibaryon in high density 
H-dibaryon matter.
In the deconfinement phase, where the Skyrmion loses its identity as the
soliton, $\bar{r}$ decreases linearly with the hypersphere radius $R$.

Finally, we compare the SO(3) soliton solution with the SO(2)$\times$SO(2) 
one, which is also a candidate of the ground state of the $B=2$ Skyrmion on a 
small hypersphere $S^3$ in the two-flavor case \cite{Jackson87}.
We show in Fig.6 the total energy as a function of the hypersphere radius $R$ 
for the SO(3) and the SO(2)$\times$SO(2) soliton solutions in the $B=2$ sector 
on $S^3$.
The energy difference is found to be quite small between the SO(3) and the 
SO(2)$\times$SO(2) solutions for the high density region as $R \lsim 1$ fm.
For instance, one finds $0.90 < E_{\rm SO(3)}/E_{{\rm SO(2)}\times{\rm SO(2)}} 
< 1.03$ for $0.2 \leq R \leq 1.0$ fm. 
One also finds $E_{\rm SO(3)} \geq 1.041 E_{\rm BPS}$ and 
$E_{{\rm SO(2)}\times{\rm SO(2)}} \geq 1.045 E_{\rm BPS}$.
From the detailed analyses, the SO(3) solution is the true ground state for 
$R \geq 0.99$ fm (dilute case) and $R \leq 0.64$ fm (dense case), while the 
SO(2)$\times$SO(2) solution is more stable for the intermediate density regions
as $0.64 \leq R \leq 0.99$ fm.
Thus, the SO(3) soliton solution is the true ground state for the small 
hypersphere with $R < 0.64$ fm in the $N_{\rm f}=3$ chiral limit in the 
mathematical sense.
In the real world, however, the large symmetry breaking in the strangeness
sector should enhance $E_{\rm SO(3)}$ largely, and therefore the 
SO(2)$\times$SO(2) Skyrmion may become more stable.

We next consider the chiral order parameter $\sigma_0$ in 
eq.(\ref{chiralorderparam}).
Since the SO(2)$\times$SO(2) solution takes the form as 
$U = \left( \begin{array}{cc}
             e^{i\taubm\cdot\pibm} & 0 \\
             0                     & 1
            \end{array} \right)$
with $\displaystyle \int_{S^3} e^{i\taubm\cdot\pibm} dV =0$ \cite{Jackson87} 
independent of $R$, $\sigma_0 = \frac{1}{3}$.
It is to be noted that the SO(2)$\times$SO(2) solution on $S^3$ provide the 
``chiral symmetric phase" only in the two-flavor sector, and does not lead to
the $N_{\rm f}=3$ chiral symmetric phase.

In summary, the SO(3) Skyrmion solution with $B=2$ has been investigated on a 
hypersphere $S^3$ to study high density H-dibaryon matter.
In this system, the deconfinement transition and the chiral symmetry 
restoration have been found at high densities similarly to $B=1$ SU(2) 
Skyrmion matter.
As a remarkable fact, the H-dibaryon in H-matter swells at high densities.
The swelling phenomenon is considered to reflect the attractive nature of the 
interaction between H-dibaryons\cite{Issinskii88}.
Such a possibility of the swelling of the H-dibaryon would be important for 
the study of H-matter in dense neutron stars, because the compactness of the 
H-dibaryon is a basic assumption for the existence of H-matter.
For the realistic treatment of H-dibaryon, we need to take account of the 
SU(3) flavor symmetry breaking.
This will be done in the future work.

\begin{figure}[t]
\caption{The profile functions, $\chi(\mu)$ and $\psi(\mu)$, of the 
$B=2$ SO(3) Skyrmion on $S^3$ vs. the angle $\mu$ of the hypersphere for 
(a) $R = 2.0$ fm, and (b) $R = 0.1$ fm.}
\label{fig1}
\begin{center}
\epsfxsize=80mm\epsfbox{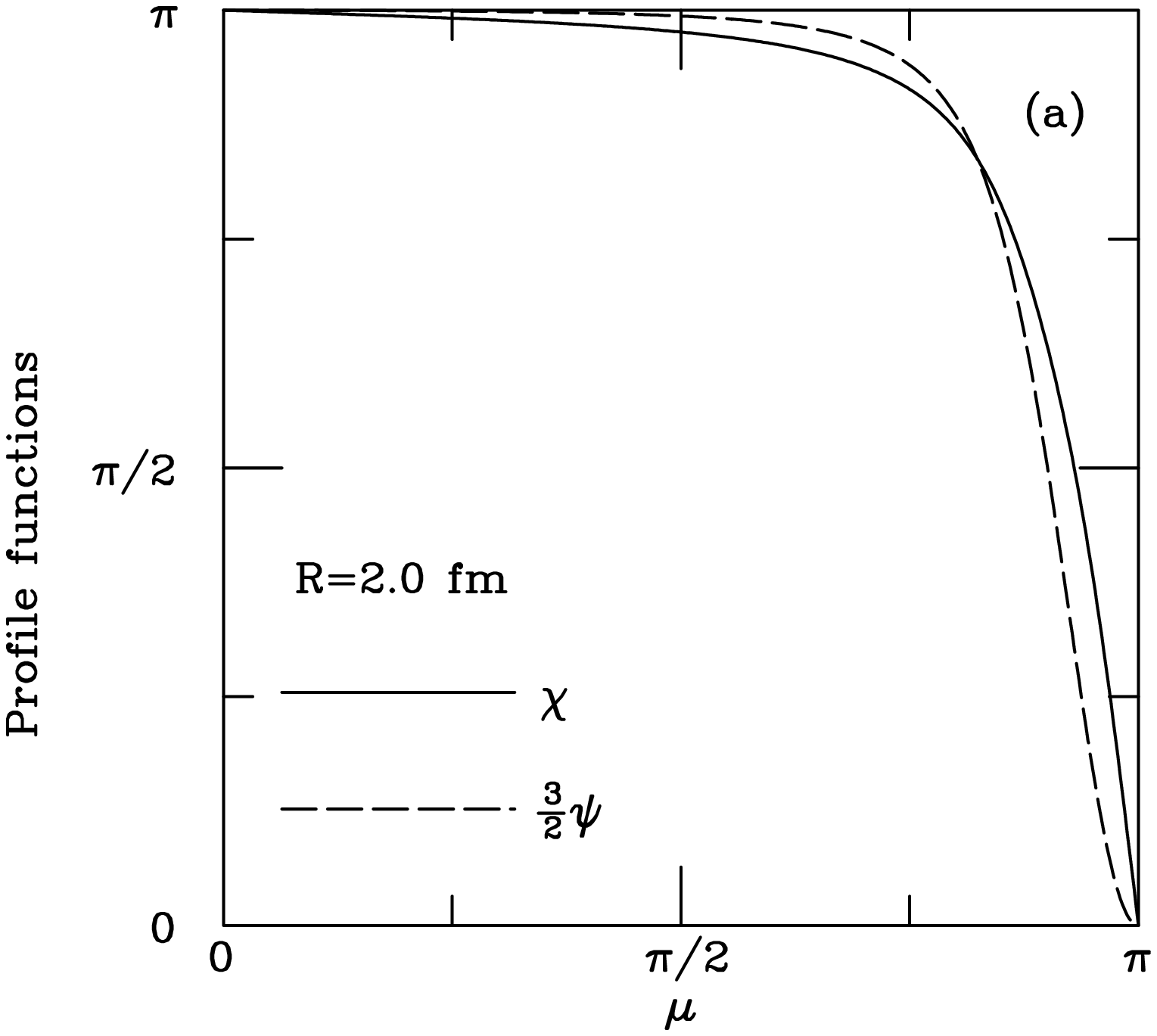}

\epsfxsize=80mm\epsfbox{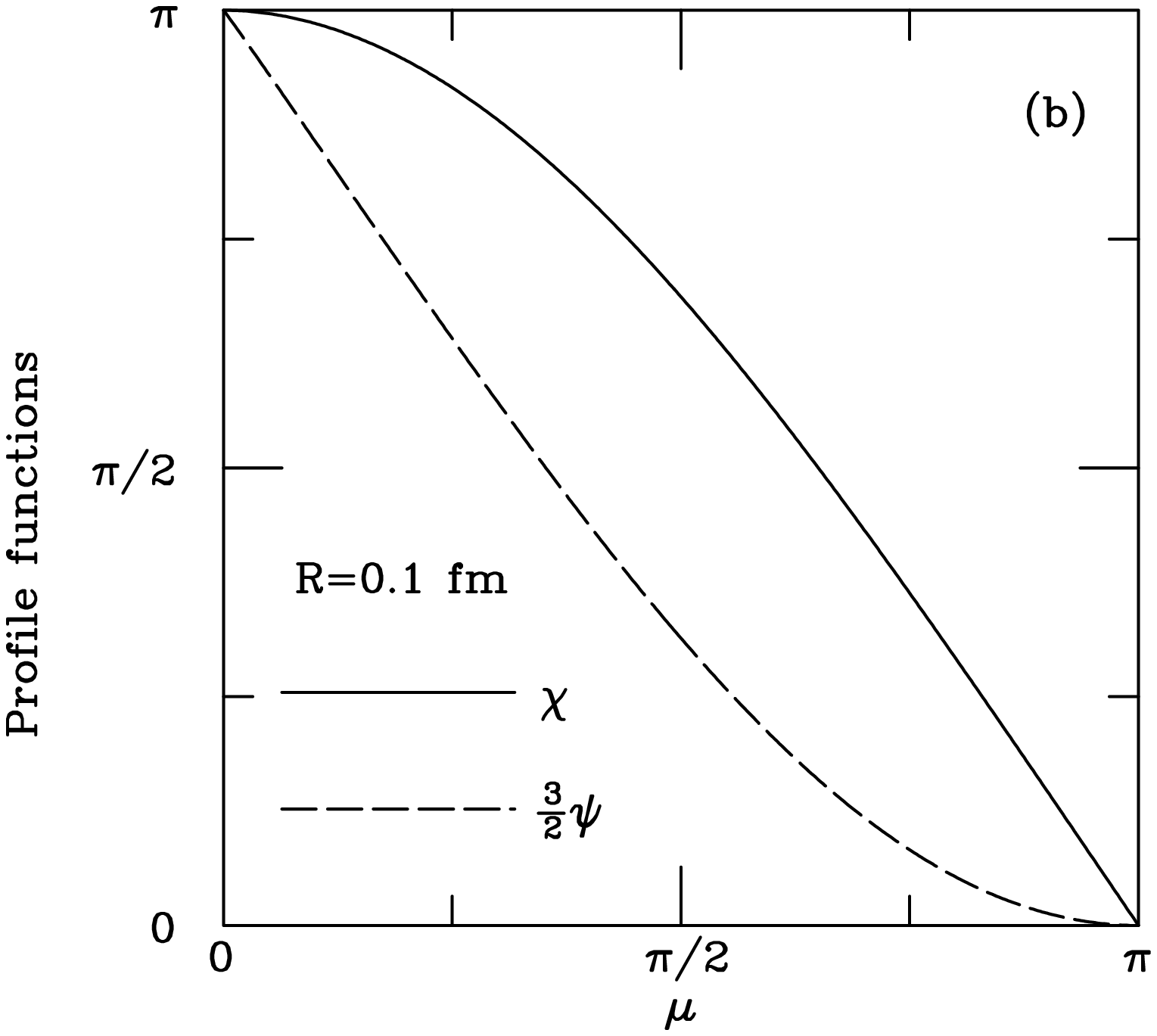}
\end{center}
\end{figure}

\begin{figure}[t]
\caption{The energy of the SO(3) Skyrmion on $S^3$ as a function of the 
hypersphere radius $R$ (solid line).
The energy in the case of the ``identity map", $\chi(\mu) = \pi - \mu$ and 
\protect{$\psi(\mu) = \frac{2}{3} (\pi - \mu)$}, is also shown by the dashed 
line.
The ``identity map" is found to be a good approximation for small $R$.}
\label{fig2}
\begin{center}
\epsfxsize=100mm
\epsfbox{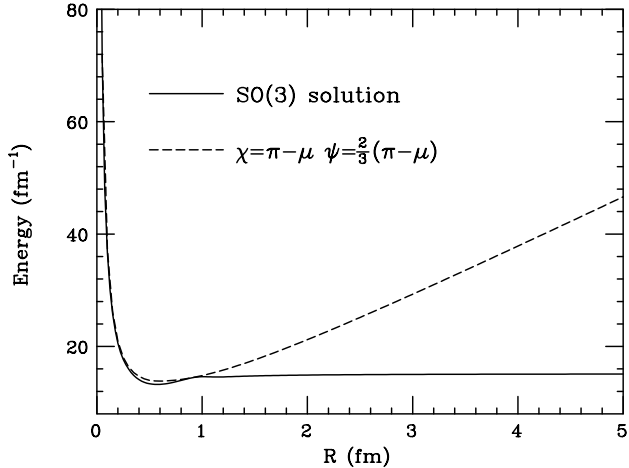}
\end{center}
\end{figure}

\begin{figure}[t]
\caption{Baryon number densities $B^0(\mu)$ vs. the angle $\mu$ of the
hypersphere for various hypersphere radii $R$.}
\label{fig3}
\begin{center}
\epsfxsize=100mm
\epsfbox{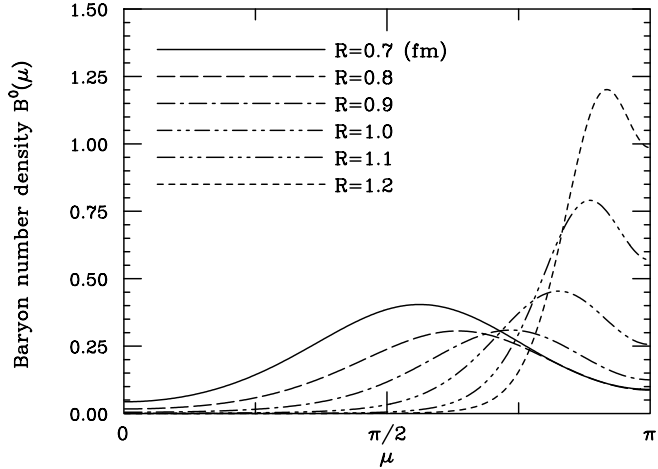}
\end{center}
\end{figure}

\begin{figure}[t]
\caption{The change of the chiral order parameter $\sigma_0$ with 
respect to the hypersphere radius $R$.}
\label{fig4}
\begin{center}
\epsfxsize=100mm
\epsfbox{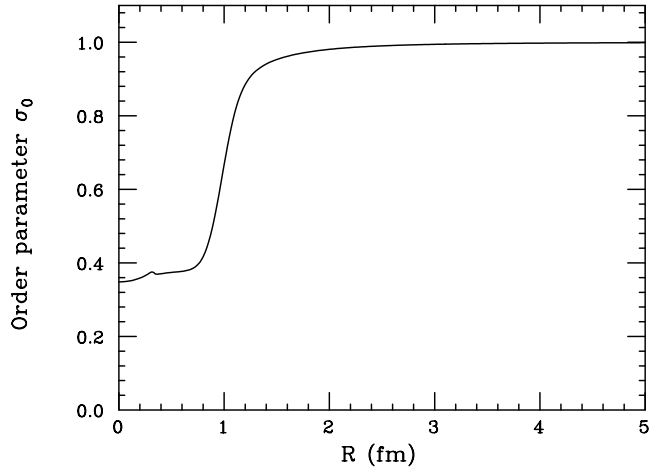}
\end{center}
\end{figure}

\begin{figure}[t]
\caption{The Skyrmion r.m.s. radius \protect{$\bar{r}$} defined by 
eq.(\protect{\ref{rmsradius}}) as a function of the hypersphere radius $R$ 
(solid line).
The contributions from the upper half ($0 \leq \mu \leq \pi /2$) and the lower 
half ($\pi /2 \leq \mu \leq \pi$) of the hypersphere are also shown by dashed 
line and dot-dashed line, respectively.}
\label{fig5}
\begin{center}
\epsfxsize=100mm
\epsfbox{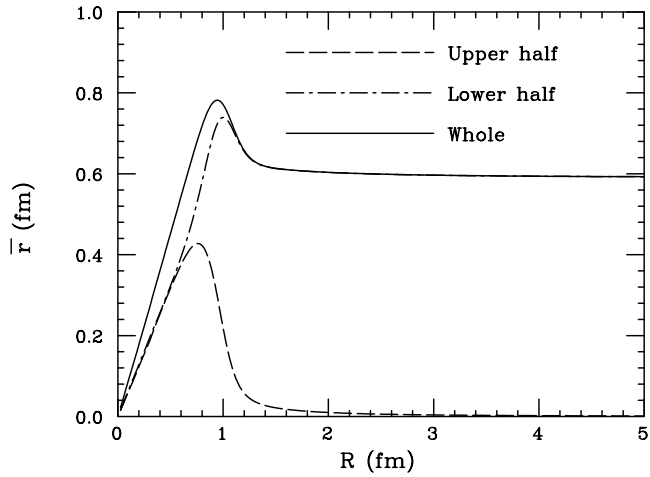}
\end{center}
\end{figure}

\begin{figure}[t]
\caption{The comparison between the SO(3) and the SO(2)$\times$SO(2) 
solutions on $S^3$.
The energies, \protect{$E_{\rm SO(3)}$} and 
\protect{$E_{{\rm SO(2)}\times{\rm SO(2)}}$}, are 
plotted as functions of the radius $R$.}
\label{fig6}
\epsfxsize=100mm
\epsfbox{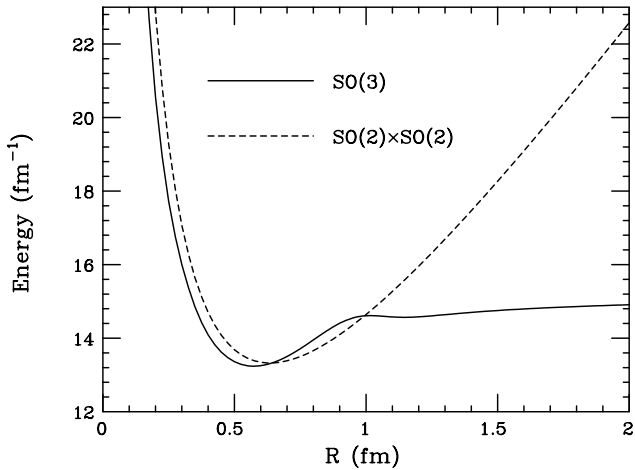}
\end{figure}

\end{document}